# Machine Learning Accelerators in 2.5D Chiplet Platforms with Silicon Photonics

Febin Sunny, Ebadollah Taheri, Mahdi Nikdast, Sudeep Pasricha
*Department of Electrical and Computer Engineering*
*Colorado State University*
Fort Collins, Colorado, USA
{febin.sunny, ebad.taheri, mahdi.nikdast, sudeep}@colostate.edu

*Abstract*—Domain-specific machine learning (ML) accelerators such as Google's TPU and Apple's Neural Engine now dominate CPUs and GPUs for energy-efficient ML processing. However, the evolution of electronic accelerators is facing fundamental limits due to the limited computation density of monolithic processing chips and the reliance on slow metallic interconnects. In this paper, we present a vision of how optical computation and communication can be integrated into 2.5D chiplet platforms to drive an entirely new class of sustainable and scalable ML hardware accelerators. We describe how cross-layer design and fabrication of optical devices, circuits, and architectures, and hardware/software codesign can help design efficient photonics-based 2.5D chiplet platforms to accelerate emerging ML workloads.

## I. INTRODUCTION

Deep neural networks (DNNs) are ubiquitously employed today in a wide range of applications, including, but not limited to, autonomous vehicles, medical diagnosis, network security, recommendation systems, and navigation solutions [1]-[5]. To cater to the objectives of these applications, DNNs have become quite varied, with the DNN family including convolution neural networks (CNNs), recurrent neural networks (RNNs), graph neural networks (GNNs), transformers, etc. A commonality among these DNN variants is the increasing model complexity and upward trend of parameter count. To meet the processing and latency demands of these applications, the hardware architecture must also scale in terms of processing capabilities, on-chip memory capacity, and on-chip communication capabilities. Graphic processing units (GPUs) are usually tasked with accelerating DNN execution today, but several limitations of the general-purpose nature of GPU architectures have become apparent in recent years. These limitations include high power consumption, increasing area overhead, reducing performance per watt, and memory bandwidth limitations [6].

The limitations of GPUs highlight the need for more efficient domain-specific accelerator architectures. However, the growing processing requirements of modern DNNs does not favor monolithic (single chip) architectures [7]. Monolithic implementations of domain specific accelerators can face scalability, power density, fabrication yield, and latency issues [8]. To tackle these problems and to effectively accelerate modern DNNs in a scalable manner, 2.5D architectures are actively being considered today [9].

Scaling 2.5D architectures comes with the challenge of increasing inter-chiplet distances. In this scenario, it can be shown that inter-chiplet metallic interconnects pose a major challenge to system performance due to excess latency and energy consumption [10]. Because of these limitations, electrical interconnect alternatives must be considered to ensure that 2.5D accelerators can deliver on the demands for low latency and energy efficient DNN acceleration.

Optical interconnects based on silicon photonics can overcome the limitations posed by metallic interconnects through advantages such as high bandwidth communication [11], single-hop data propagation [12], and high energy efficiency [13]. Silicon photonic interconnects also allow for ease of broadcast [13], [14], which is a desirable feature for DNN acceleration [15]-[17]. Further energy and latency benefits can be extracted from photonics by utilizing photonics for computation as well. Many prior efforts have shown that photonic processing for DNN inference acceleration provides significant benefits in terms of latency and energy efficiency [18]-[26]. Thus, it stands to reason that utilizing photonics for both communication and computation may amplify the aforementioned benefits. In this work, we explore the benefits provided by silicon photonic chiplets and networks for DNN acceleration in 2.5D chiplet platforms.

The organization of the remainder of this paper is as follows. An overview of silicon photonics is provided in Section II. Section III discusses silicon-photonic based DNN accelerators. In Section IV, state-of-the-art silicon photonic interposer networks are presented. Section V describes our 2.5D chiplet-based DNN accelerator. Section VI presents various experimental results. Finally, Section VII provides conclusions and open challenges in this emerging area.

## II. SILICON PHOTONICS OVERVIEW

Silicon photonics emerged as a CMOS-compatible technology to enable chip-scale optical communication. To achieve this, silicon-on-insulator (SOI) waveguides are employed, which use silicon (Si) for the core material and silicon-dioxide ($SiO_2$) for cladding and substrate material. Moreover, by using wavelength-division multiplexing (WDM), optical signals on different wavelengths can simultaneously traverse the same waveguide. Silicon photonics promises high energy efficiency, bandwidth density, and low latency, as the overall scale in terms of communication distances increases [27]. Due to the benefits that silicon photonics offers, there has been a growing interest in using silicon photonics for computation, including realizing digital logic using photonics [28], [29]. To realize any photonic-based computation or network system, there is a need for many fundamental components, as discussed next.

Silicon photonic waveguides are analogous to metallic wires in electrical chips and enable optical signal transmission and routing in chips. Photonic waveguides operate on the principle of total internal reflection (TIR) to contain and guide optical signals [30]. To ensure TIR, these waveguides require high refractive index contrast between their core and cladding materials (e.g., an SOI platform).

Lasers are a key requirement for any photonic system as they act as light sources for communication and computation. The laser sources employed can be on-chip or off-chip [31]. Off-chip lasers offer better light emission efficiency, but they face high optical power losses during coupling to on-chip waveguides. On-chip lasers provide better integration density and lower optical loss, as there is no need to couple light, but they suffer from low light emission efficiencies. In most photonics-based systems, the laser that is commonly utilized is a vertical cavity surface emission laser (VCSEL) or a microring laser. If off-chip lasers are used, then couplers are necessary devices to couple the optical signals from off-chip sources to on-chip waveguides. Coupling solutions employed can be surface grating couplers or edge couplers [33].

Microring resonators (MRs) are photonic devices that are widely used to design modulators, switches, and optical filters [34]. In computation systems, they can be used to perform multiplication operations, through amplitude modulation [35]. MRs are fabricated with a ring-shaped silicon photonic waveguide (see Fig. 1). An MR can be in one of two different states of on- or off-resonance, based on which the optical signal can be switched to different ports. The resonant wavelength of an MR can be tuned using electro-optic (EO) or thermo-optic (TO) effects of silicon that alter the effective index of the composite waveguide of the device. In a communication system, active MRs or MRs with a tuning circuit are used to filter wavelengths that correspond to 0s in an on-off keying (OOK) modulation scheme. In advanced modulation schemes such as 4 pulse amplitude modulation (PAM-4) [44], MRs can be used to modulate signal amplitude on four distinct levels. Multiple MRs sensitive to the same wavelength can be used for consecutive amplitude modulation resulting in parameter multiplication. Microdisk resonators [36] are similar to MRs but are composed of a disk structure instead of a ring structure. They are more compact than MRs but have higher operation losses.

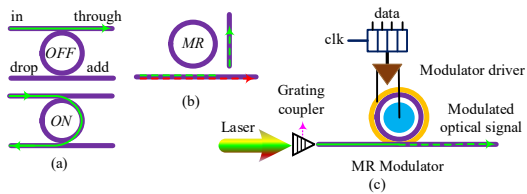

*Fig.1 Microring resonator (MR): (a) off / on states, (b) MR filter, and (c) MR modulator.*

Mach-Zehnder interferometers (MZIs), are made of two 3-dB directional couplers and two waveguide arms with phase shifters. The phase shifters, implemented using electro-optic or thermo-optic tuning, can change the optical phase in one or both arms of the MZI, introducing constructive or destructive interference at the output, to switch an optical signal between the output ports. MZIs are applied to the design of optical modulators, switches, and filters. MRs have a smaller footprint and lower power consumption than MZIs. However, MZIs provide better thermal stability in operation and better extinction ratios than MRs [20].

Photodetectors (PD) are used to convert photonic signals to electrical signals. The operation of a PD may trade off bandwidth of operation with power efficiency. An efficient PD provides the desired electrical output with a small optical signal at its input. However, this small optical signal at the input of a PD may result in a low-bandwidth performance. For an efficient conversion of a photonic signal to an electrical signal, the intensity of the photonic signal received by the PD should be larger than the responsivity of the PD. High-bandwidth PDs can be employed in photonic computation to perform accumulation operations across signals of different wavelengths [32].

III. SILICON-PHOTONIC-BASED DNN ACCELERATORS

With the promise of improved energy efficiency and latency, silicon photonics for DNN acceleration has become increasingly prominent in both academic and industrial research [36]. Photonics is especially suited to accelerate DNN inference operations, which rely heavily on fixed matrix multiplications. The linear transformations involved in matrix multiplication can be implemented efficiently in the analog domain using photonics. Silicon photonic DNN accelerators can be implemented as either coherent or non-coherent architectures, as described below.

Coherent architectures utilize a single wavelength and rely on constructive and destructive interference to change the relative power levels of a coherent optical beam [18]-[20]. Optical phase control is used to imprint the parameters onto the light wave signals. To achieve this, coherent architectures make use of MZIs, with phase modulators embedded on their arms. Weighting occurs with electrical field amplitude attenuation proportional to the weight value, and phase modulation that is proportional to the sign of the weight. Cascaded combiners, which facilitate coherent interaction of the signals, are used for accumulation. A lot of work in this field is focused on reducing the computational complexity of the DNN being implemented on-chip. The reduction of computational complexity is achieved using pruning methods [18] or singular value decomposition (SVD) [19], [20].

Noncoherent architectures, such as [21]-[26], use multiple wavelengths, where each wavelength can be used to perform computations in parallel. In these architectures, parameters are imprinted onto the signal amplitude using wavelength-selective devices, such as MRs. Several prior works, as mentioned above, have explored DNN acceleration using non-coherent photonic principles. In [21], an MR-based DNN accelerator architecture was proposed which utilizes modular vector-dot-product units with optimized MR designs and tuning circuit optimization, for energy and throughput efficiency. For further optimizing power and energy consumption of non-coherent accelerators, especially at the electrical-photonic interface, [22] employed heterogeneous quantization (i.e., potentially different parameter bit-widths for each DNN layer) along with hardware-software co-optimization. For lowering area and power consumption, the work in [23] utilized microdisks instead of MRs. For further reducing the power consumption at the electro-optical interface, binarized neural networks can be considered. A microdisk-based photonic accelerator was proposed in [24] for fully binarized DNNs (single-bit weight and activation parameters). While fully binarized neural networks offer higher efficiency in storage and power consumption, they may lack in achievable accuracy. To tackle the accelerator needs of partially binarized neural networks, the work in [25] proposed an MR-based partially binarized DNN accelerator. Non-coherent architectures have also been proven effective for RNN acceleration, as shown in [26]. In [26], the speed of operation of the photonic accelerator substrate was used to perform large-scale matrix operations needed for different types of RNNs, including deep long short-term memory (LSTM) and gated recurrent unit (GRU) models.

## IV. SILICON PHOTONIC INTERPOSER NETWORKS

Conventionally, chiplet systems are packaged using passive [39] and active [40] electronic interposers. Compared to passive electronic interposers, active electronic interposers employ an interconnection fabric with logic elements, instead of only passive metal interconnects to offer better communication scalability. However, both active and passive electronic interposers are unable to efficiently support a system with a large number of chiplets due to the inherent limitations of metallic interconnects: high latency for long interconnects and low bandwidth per each interconnect.

On the other hand, as optical interconnects offer low latency and high bandwidth, a photonic interconnection fabric can be a promising solution for interposer designs. Therefore, photonic interposers have recently received much attention in chiplet systems [10], [11]. For example, [10] employs high-bandwidth arrayed-waveguide grating routers (AWGRs) to get around the high latency and low bandwidth of conventional electronic interposers used in chiplet systems. Besides the high bandwidth and low latency in long interconnects, silicon photonic interposers are inherently capable of dynamic inter-chiplet bandwidth tuning. PROWAVES [11] describes a photonic interposer network that dynamically manages inter-chiplet bandwidth by tuning the number of active wavelengths with respect to the traffic load. Under a low traffic load, where low bandwidth is required, PROWAVES utilizes a smaller number of wavelengths and deactivates unused wavelengths in an off-chip laser to save power consumption. On the other hand, under a high traffic load, PROWAVES activates a larger number of wavelengths to offer a high bandwidth and, as a result, high performance for inter-chiplet communication at the cost of higher power consumption. To handle high traffic load in PROWAVES, a high bandwidth gateway on each chiplet is used (i.e., a gateway with a large number of wavelengths). However, in this architecture, the high bandwidth gateway can create congestion on the chiplet as all the nodes on the same chiplet utilize the same gateway to communicate with the interposer. Moreover, access to the gateway is not enabled in a fair manner for the nodes across the chiplets. For example, a node far from the gateway can encounter very high latency to reach the gateway.

ReSiPI [37] improves on the PROWAVES design by employing several gateways on a chiplet with a relatively lower number of wavelengths. Moreover, ReSiPI manages inter-chiplet bandwidth while considering the online traffic by tuning the number of active gateways instead of the number of active wavelengths. In the ReSiPI architecture, the traffic load of inter-chiplet communication is monitored in time epochs and the number of active gateways is defined based on the required inter-chiplet bandwidth. Activating or deactivating gateways is done using a phase-change-material-based coupler (PCMC), based on the coupler design in [38]. As shown in Fig. 2, a PCMC can be in three states: 1) crystalline state to guide input light to the Bar (B) output, 2) partially crystalline state to guide a portion of input light to the Cross (C) output and the rest to the Bar output, and 3) amorphous state to guide the light to the Cross output. $CL_{am}$ and $CL_{cr}$ are the coupling lengths of the amorphous and crystalline states, respectively. By tuning the ratio of $CL_{am}$ to $CL_{cr}$ the appropriate input optical power from an optical laser to a writer gateway can be adjusted. Typically, in a silicon photonic network where a writer gateway modulates data on an optical signal to be received by a reader gateway, passive splitters are used to divide and deliver the optical signal from the optical laser to the writers. However, passive splitters prevent the network from dynamically deactivating writer gateways, while the PCMC can tune optical input of each writer and facilitate dynamic gateway activation and deactivation. Using the PCMC, the ReSiPI interposer is designed to reconfigure the number of active gateways and improve power consumption of the network. A controller is used to tune the number of active gateways in each chiplet according to the inter-chiplet traffic of that chiplet. Based on the number of active gateways, the PCMCs are tuned to deliver appropriate optical power to each gateway. Besides tuning the PCMC, the controller also tunes the laser power accordingly, to save the power consumption of the laser.

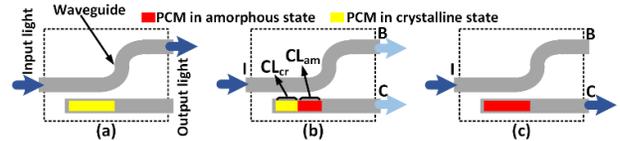

*Fig. 2 PCM-based coupler (PCMC) used in ReSiPI with three states: (a) crystalline, (b) partially crystalline, and (c) amorphous.*

## V. SILICON PHOTONIC 2.5D DNN ACCELERATORS

To explore the implications of accelerating DNNs on 2.5D interposer platforms, we present a case study that involves extending the CrossLight [21] photonic DNN accelerator to the 2.5D chiplet platform.

CrossLight is a neural network accelerator designed to perform high speed multiply and accumulate (MAC) operations in the photonic domain. However, the original monolithic CrossLight architecture suffers from low scalability and relatively low energy efficiency. We propose to use a ReSiPI-based photonic interposer architecture to design a more scalable and energy-efficient 2.5D CrossLight implementation. A high-level overview of our chiplet-based 2.5D CrossLight accelerator with a photonic interposer is shown in Fig. 3. Several chiplets are packaged on a silicon photonic interposer substrate. We consider different types of chiplets as part of a heterogeneous architecture. Such a heterogeneous design allows system-on-chip (SoC) designers to utilize appropriate off-the-shelf chiplets and create diverse 2.5D packages to meet their design targets [8].

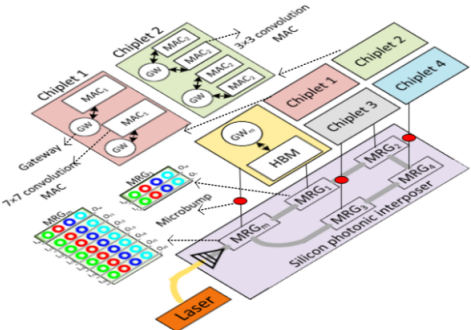

*Fig. 3 Overview of proposed 2.5D interposer chiplet-based DNN accelerator architecture.*

The chiplets in the proposed architecture consist of various computational and memory chiplets. One or more chiplets consist of an optically-interfaced memory architecture, such as high bandwidth memory (HBM; shown in Fig. 3), with a dedicated gateway to communicate with the rest of the system. Each compute chiplet (e.g., chiplets 1-4 in Fig. 3) hosts several photonic MAC units and has its local gateway(s) to read data from the memory chiplets and write data to them through the interposer network. Each gateway has two main parts: electronic circuitry on the chiplet and a Microring Resonator Group (MRG) on the interposer. The electronic part of a gateway is connected to the microrings of an MRG using the microbump technology.

The photonic MAC units utilize noncoherent photonics to perform multiply operations between parameters, and photodetectors are used to obtain the sum of products. The weights and activations are imprinted on wavelengths using banks of wavelength-specific MR filters. The imprinting process follows the broadcast-and-weight protocol as described in [35]. Even though the proposed design utilizes photonic communication to move data, moving multi-bit amplitude modulated data in a robust manner is challenging. Thus, the interposer relies on on-off keying (OOK) based data transmission, with intermediate photonic-to-electronic conversion at the gateways and buffering of the parameters at the MAC units. The buffered data is used to tune the respective MRs so that the parameter value can be represented using the wavelength amplitude. For tuning the MRs, EO tuning is used. The proposed architecture employs heterogeneous MAC unit sizes (size referring to the size of the vectors that can be deployed) across different chiplets to cater to the different kernel sizes and to handle the large-scale MAC operations needed for the fully connected layers. For example, in Fig.3, Chiplet 1 includes 3×3 convolution MACs, while Chiplet 2 contains 7×7 convolution MACs. Moreover, as footprint of MACs with various sizes are different, the number of MACs per chiplet can vary for each chiplet. Fig. 4 shows an overview of a MAC unit.

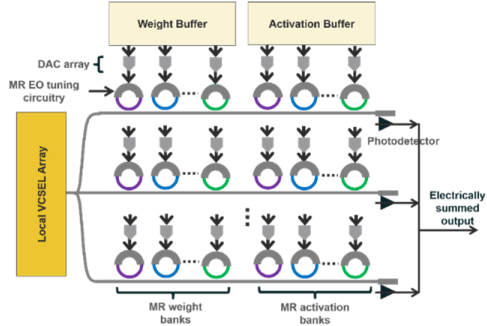

*Fig. 4 MAC unit architecture (DAC: digital to analog converter)*

An example of the optical interface and communication on the interposer in this architecture is shown in Fig. 5. In this example, MACs are reading data from the HBM on a separate chiplet. For successful communication, a writer gateway, including buffers to store and forward data, is utilized in the HBM chiplet, and similarly, a reader gateway is utilized on the chiplet with MAC units. The stored data in the buffers of the writer gateway is modulated on the optical signals which are generated by an off-chip laser. Different colors of modulators show that they are used to modulate different optical signals on different wavelengths. As discussed earlier, employing several optical signals with different wavelengths enables our network to transmit more data at the same time on the same waveguide, to improve the communication bandwidth. Several MR filters are also connected to the reader gateways. Each MR filter is tuned at a specific wavelength to filter and drop the specified optical signal. After this step, the optical signal is converted to an electronic signal using a photodiode, and this signal is delivered to the reader chiplet using microbumps. The reader gateway converts the electronic signal to digital data, and stores the received data in its buffer. Finally, the data will be forwarded to the MACs. Such a protocol for optical communication, where a reader is receiving data from a writer using a waveguide, is called the single writer single reader (SWSR) protocol. Similarly, if several readers are receiving data from a writer using a waveguide, the protocol is referred to as single writer multiple reader (SWMR) protocol.

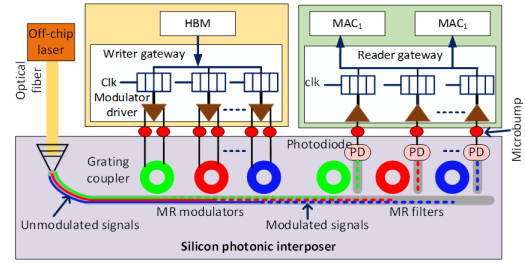

*Fig. 5 Example of optical communication on interposer: MACs are reading data from memory.*

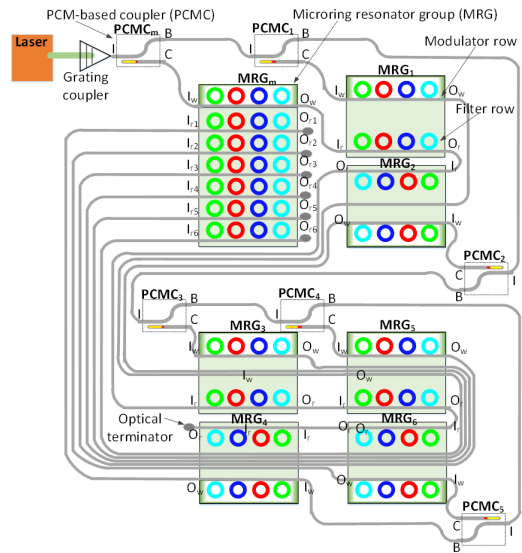

*Fig. 6 Silicon photonic network in our 2.5D chiplet-based DNN accelerator. Each MRG is connected to a gateway on a chiplet.*

In our architecture, we have two types of traffic between the chiplets: 1) reading weights and inputs needed by MACs from memory, and 2) writing MAC outputs to the memory. As a result, from the memory chiplet to the compute chiplets, we utilize the SWMR protocol to perform reads from memory. Moreover, from the compute chiplets to the memory, we use the SWSR protocol. Therefore, the MRG of the memory chiplets requires several sets of MR filters (each set of MR filters is a row of the MRG shown in Fig. 3) to receive data from the compute chiplets. On the other hand, a compute chiplet requires only one set of MR filters as it only receives data from the memory. Both compute and memory chiplets require one set of MR modulators to send data.

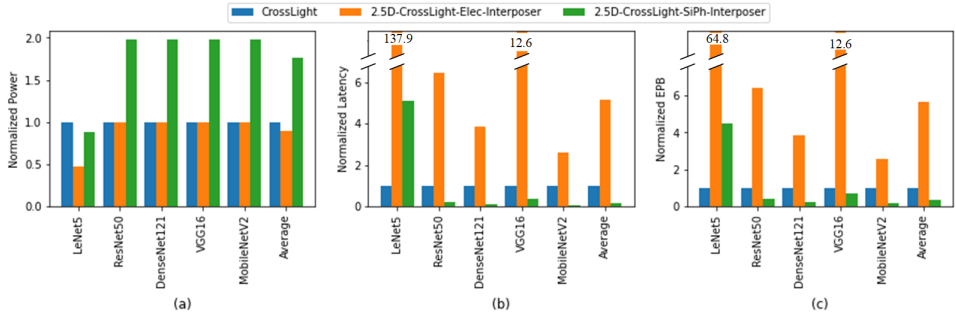

*Fig. 7 Performance analysis of CrossLight, 2.5D-CrossLight with electronic interposer, and 2.5D-CrossLight with silicon photonic interposer, (a) normalized power consumption, (b) normalized total latency, and (c) normalized energy-per-bit*

An example of our 2.5D CrossLight with the integrated ReSiPI interposer is shown in Fig. 6. Although this example is shown with six gateways (associated with one memory chiplet and five compute chiplets), the interposer design can be extended to a larger system without loss of generality. As shown in Fig. 6, the MRG of the memory chiplet ($MRG_m$) has six filter rows to receive data from the six gateways of the compute chiplets ($MRG_1 – MRG_6$), while $MRG_m$ has one row of modulators to send data to all the gateways. The photonic interposer network architecture is a passive network to save energy. This means that there is a specific waveguide to transmit data from each writer gateway to each reader gateway and the route (waveguide) does not change.

TABLE 1. MODELING PARAMETERS

| Parameter | | Value |
|---|---|---|
| Data rate of optical link (per wavelength) | | 12 Gb/s |
| Gateway frequency | | 2 GHz |
| Electrical network-on-chip link width | | 128 bits |
| Electrical network-on-chip frequency | | 2 GHz |
| Number of wavelengths | | 64 |
| Number of memory-chiplets | | 1 |
| Number of compute-chiplets | | 8 |
| 100 unit dense MAC | Number of chiplets | 2 |
| | Number of MACs per chiplet | 4 |
| | Number of MACs per gateway | 1 |
| 7×7 convolution MAC | Number of chiplets | 1 |
| | Number of MACs per chiplet | 8 |
| | Number of MACs per gateway | 2 |
| 5×5 convolution MAC | Number of chiplets | 2 |
| | Number of MACs per chiplet | 16 |
| | Number of MACs per gateway | 4 |
| 3×3 convolution MAC | Number of chiplets | 3 |
| | Number of MACs per chiplet | 44 |
| | Number of MACs per gateway | 11 |

TABLE 2. CONSIDERED DNN MODELS IN OUR EVALUATION.

| Model | CONV layers | FC layers | Parameters |
|---|---|---|---|
| LeNet5 | 3 | 2 | 62,006 |
| ResNet50 | 53 | 1 | 25,636,712 |
| DenseNet121 | 120 | 1 | 8,062,504 |
| VGG16 | 13 | 3 | 138,357,544 |
| MobileNetV2 | 52 | 1 | 3,538,984 |

## VI. EXPERIMENTAL RESULTS

We designed two variants of the 2.5D CrossLight architecture: with a ReSiPI-based interposer [37] (*2.5D-CrossLight-SiPh-Interposer*), and an electrical mesh interposer [40] (*2.5D-CrossLight-Elec-Interposer*). We also compare the two 2.5D CrossLight variants with the original monolithic (single-chip) CrossLight architecture in terms of power, latency, and energy efficiency. The model parameters assumed in this study are summarized in Table 1. We also employ the power model and power parameters used in [11] and [37]. We consider one memory chiplet and eight compute chiplets in which two of the chiplets include dense-layer MACs and six of them include convolution layer MACs (3×3, 5×5 and 7×7 convolution MACs). We used various DNN models, summarized in Table 2, for our evaluation.

The performance results are shown in Fig. 7. In general, *2.5D-CrossLight-SiPh-Interposer* is able to achieve superior energy efficiency and latency across almost all models, except for very small ones (e.g., LeNet5). The heterogeneous chiplets and high bandwidth inter-chiplet photonic network enable more energy-efficient execution of DNNs than in the monolithic *CrossLight* case.

*2.5D-CrossLight-SiPh-Interposer* imposes a non-trivial power overhead as its photonic network consumes higher power for communication than an electronic network. However, *2.5D-CrossLight-SiPh-Interposer* has relatively lower power consumption for smaller DNN models (e.g., LeNet5) as the ReSiPI controller reconfigures the photonic interposer and deactivates unnecessary gateways. Nonetheless, for the smaller model (LeNet5), where each layer only takes up a small fraction of the overall compute real estate, the *2.5D-CrossLight-SiPh-Interposer* overheads become significant and adversely affect energy efficiency.

For larger models where multiple layers are mapped to chiplets, the *2.5D-CrossLight-SiPh-Interposer* overheads in terms of power consumption are better amortized across these mappings. The controller also activates gateways in the large models to cope with high traffic volumes, which helps to improve inter-chiplet latency. Although *2.5D-CrossLight-Elec-Interposer* has lower power consumption, it suffers due to the significantly higher latency of metallic interconnects, especially for relatively long distances on large interposers.

TABLE 3. AVERAGE POWER, LATENCY, AND ENERGY-PER-BIT ACROSS ELECTRONIC AND PHOTONIC DNN ACCELERATOR PLATFORMS.

| | Power (W) | Latency (ms) | EPB (nJ/bit) |
|---|---|---|---|
| CrossLight [21] | 50.8 | 8 | 3.6 |
| 2.5D-CrossLight-Elec | 45.3 | 41.4 | 20.5 |
| 2.5D-CrossLight-SiPh | 89.7 | 1.21 | 1.3 |
| Nvidia P100 GPU | 250 | 13.1 | 12.3 |
| Intel 9282 CPU | 400 | 86.5 | 64.4 |
| AMD 3970 CPU | 280 | 141.3 | 73.7 |
| Edge TPU | 2 | 2366.4 | 17.6 |
| Null Hop [42] | 2.3 | 8049.3 | 68.9 |
| Deap_CNN [43] | 122 | 619.01 | 1959.4 |
| HolyLight [23] | 66.5 | 86.4 | 40.3 |

On average, in comparison with monolithic *CrossLight*, *2.5D-CrossLight-SiPh-Interposer* shows 6.6× lower latency, which also results in 2.8× lower energy-per-bit (EPB). Compared to *2.5D-CrossLight-Elec-Interposer*, *2.5D-CrossLight-SiPh-Interposer* offers 34× lower latency and

15.8× lower EPB. Such significant improvement comes from the ability in *2.5D-CrossLight-SiPh-Interposer* to select appropriate chiplets to map layers of each DNN model and tuning the required inter-chiplet bandwidth accordingly. We also compared *2.5D-CrossLight-SiPh-Interposer* accelerator with state-of-the-art accelerators in terms of average power, latency (total latency of layers), and EPB (Table 3). As *2.5D-CrossLight-SiPh-Interposer* performs well for larger models and also outperforms state-of-the-art electronic and photonic accelerators (in terms of latency and EPB), such a photonics-based 2.5D chiplet platform shows great promise to support acceleration of emerging large DNN models.

## VII. Conclusions and Open Challenges

In this paper, we presented a 2.5D chiplet platform-based photonic DNN accelerator where both communication on the interposer and computation on the chiplets employ silicon photonics. Compared to a monolithic photonic accelerator, a chiplet-based one not only improves fabrication yield and cost, but also reduces latency using a high-bandwidth photonic network on the interposer. Moreover, chiplets can be designed heterogeneously and off-the-shelf chiplets can be integrated in 2.5D packages to make various systems with different computation power budgets and capabilities.

There are several open challenges in this field to design a more efficient silicon photonic DNN accelerator: 1) power consumption of the state-of-the-art photonic devices are relatively high, and there is a need for device-level efforts to design low-power devices; 2) designing an efficient electronic controller is essential to efficiently control the communication and computation operations with low latency; and 3) the silicon photonic 2.5D DNN accelerator architecture requires design-space exploration (e.g., in terms of the number of wavelengths, number of gateways per chiplet, and number of MACs per chiplet) to create an optimized architecture tailored to DNNs of interest.